\documentclass[9pt,twocolumn,twoside]{osajnl}
\usepackage{graphicx} % figure lookup in several defined folders
\graphicspath{{Figures/}}

\usepackage{subfig}
\usepackage{caption}
\journal{ol} % Choose journal (ao, aop, josaa, josab, ol, pr)

\setboolean{shortarticle}{true}
\title{Carrier-envelope offset stable, coherently combined ytterbium-doped fiber CPA delivering 1\,kW of average power}

\author[1,2*]{E. Shestaev}
\author[1]{S. Hädrich}
\author[1]{N. Walther}
\author[1]{T. Eidam}
\author[2,3]{A. Klenke}
\author[4]{I. Seres}
\author[4]{Z. Bengery}
\author[4]{P. Jójárt}
\author[4]{Z. Várallyay}
\author[4]{Á. Börzsönyi}
\author[1,2,3,5]{J. Limpert}

\affil[1]{Active Fiber Systems GmbH, Ernst-Ruska-Ring 17, 07745 Jena, Germany}
\affil[2]{Friedrich-Schiller-Universität Jena, Abbe Center of Photonics, Institute of Applied Physics, Albert-Einstein-Straße 15, 07745 Jena, Germany}
\affil[3]{Helmholtz-Institute Jena, Fröbelstieg 3, 07743 Jena, Germany}
\affil[4]{ELI-ALPS, ELI-HU Non-Profit Ltd., Wolfgang Sandner utca 3., Szeged, H-6728, Hungary}
\affil[5]{Fraunhofer Institute for Applied Optics and Precision Engineering IOF, Albert-Einstein-Straße 7, 07745 Jena, Germany}

\affil[*]{Corresponding author: evgeny.shestaev@uni-jena.de}

\begin{abstract}
We present a carrier-envelope offset (CEO) stable ytterbium-doped fiber chirped-pulse amplification system employing the technology of coherent beam combining and delivering more than 1\,kW of average power at a pulse repetition rate of 80\,MHz. The CEO stability of the system is 220\,mrad rms, characterized out-of-loop with an f-to-2f interferometer in a frequency offset range of 10\,Hz to 20\,MHz. The high power amplification system boosts the average power of the CEO stable oscillator by five orders of magnitude, while increasing the phase noise by only 100\,mrad. No evidence of the CEO noise deterioration due to coherent beam combining has been found. Low-frequency CEO fluctuations at the CPA are suppressed by a "slow loop" feedback. This is the first demonstration of a coherently combined laser system delivering an outstanding average power and a high CEO stability at the same time.
\end{abstract}

\setboolean{displaycopyright}{true}

\emergencystretch=\maxdimen
\begin{document}

\maketitle

%INTRODUCTION

Carrier-envelope phase (CEP) stable few-cycle sources have revolutionized science by allowing for experiments revealing the dynamics of electrons~\cite{calegari_ultrafast_2014}, molecules ~\cite{niikura_sub-laser-cycle_2002} and chemical bonds~\cite{zewail_femtochemistry_2000}, finally leading to a development of an entirely new research area - attosecond science~\cite{corkum_attosecond_2007}. Currently developed next-generation femtosecond light sources are intended to produce few-cycle pulses at a high repetition rate and average power. As noted by a number of researchers, future applications in e.g. high-harmonic generation~\cite{elouga_bom_attosecond_2011}, attosecond streaking~\cite{boltaev_high-order_2020} and precision spectroscopy~\cite{wernet_orbital-specific_2015} would greatly benefit from a higher photon flux, which would significantly reduce the experiment duration or improve the signal-to-noise ratio.

A demand for a CEP-stable laser source delivering 5\,mJ, 6\,fs pulses at a repetition rate of 100\,kHz and an average power of 500\,W has been stated by the ELI-ALPS research facility in Szeged, Hungary, whose primary mission is ``to provide the international scientific community with attosecond sources beyond the current state of the art in terms of repetition rate, intensity and reliability''~\cite{kuhn_eli-alps_2017}. One solution to this challenge is employing an ytterbium-doped fiber chirped-pulse amplifier (CPA) followed by a hollow-core fiber nonlinear pulse compressor. In our previous work, a laser system following this strategy and delivering 100\,\textmu J, sub-8\,fs pulses at a repetition rate of 100\,kHz with a CEP stability of 360\,mrad rms (10\,Hz to 50\,kHz) has been demonstrated~\cite{shestaev_high-power_2020}.

To date, the highest average power achieved by a highly CEO stable laser system (265\,mrad rms) is 132\,W~\cite{luo_130_2020}. Further scaling of the laser parameters is allowed for by using the technique of coherent beam combining~\cite{klenke_coherent_2018}. However, the CEP noise properties of a coherently combined CPA system have yet to be demonstrated, which is the focus of the present work.

% MAIN
%In this Letter, we present a coherently combined chirped-pulse ytterbium-doped fiber amplifier delivering more than 1\,kW of average power at a repetition rate of 80\,MHz with an excellent carrier-envelope offset stability of 220\,mrad rms measured in a broad offset frequency range of 10\,Hz to 20\,MHz. This is the first demonstration of a highly CEO-stable coherently combined kilowatt-level fiber laser CPA system.

The reported system (Fig.~\ref{fig:setup}) represents a CPA of the ELI-ALPS HR2 laser system {\cite{kuhn_eli-alps_2017}, which is currently developed by Active Fiber Systems GmbH and targeting generation of 5\,mJ few-cycle, CEP stable pulses at a repetition rate of 100\,kHz. The system is seeded by a commercial CEO-stable fiber oscillator operating at a pulse repetition rate $f_\text{rep}$ of 80\,MHz with a CEO lock frequency $f_\text{ceo} = f_\text{rep}/4 = 20$\,MHz. 

%FIG1: SETUP
\begin{figure}[]%[htbp]
\centering
\includegraphics[width=\linewidth]{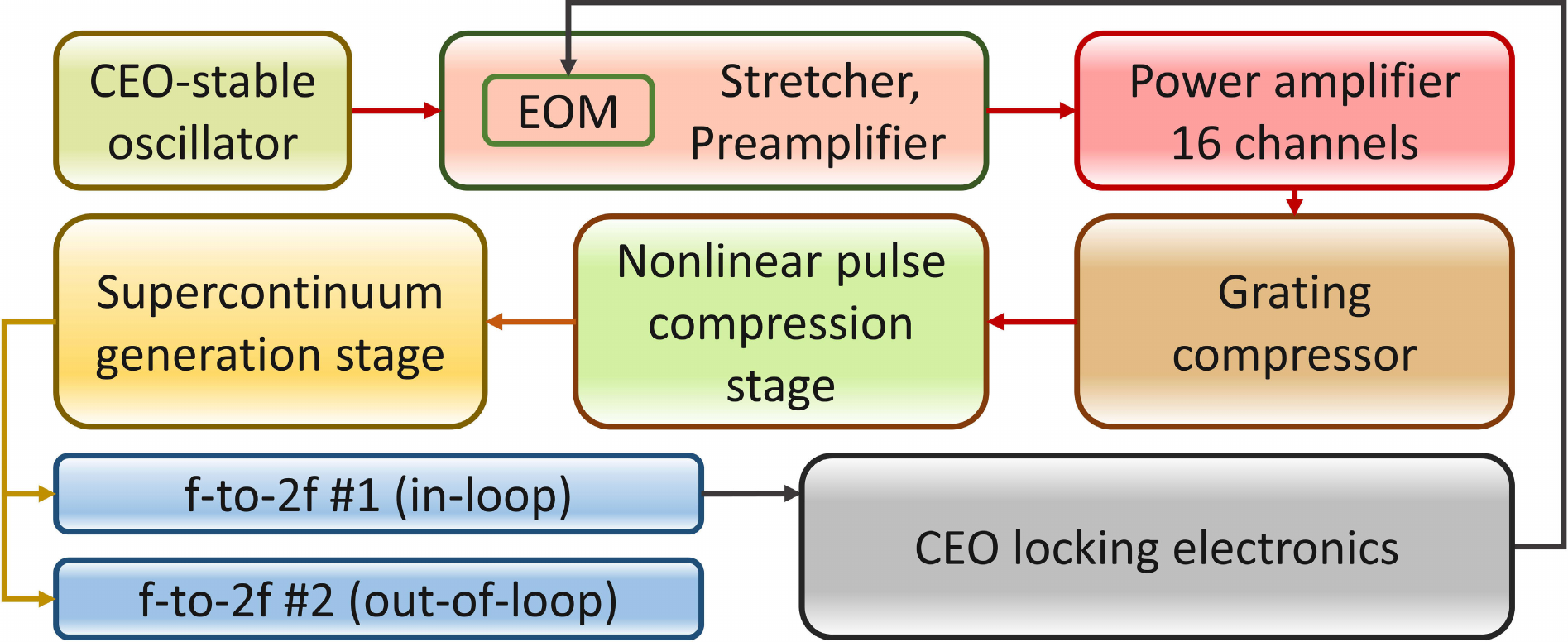}
\caption{Experimental setup: a coherently combined, fiber CPA system is seeded by a CEO-stable fiber oscillator. The CEO noise is measured with two f-to-2f interferometers with shared nonlinear compression and supercontinuum generation stages. A "slow loop" feedback is performed to an EOM in the front-end.}
\label{fig:setup}
\end{figure}

The pulses are stretched to 3.8\,ns, amplified and spectrally shaped in an all-fiber front-end based on polarization-maintaining components, which ensures high environmental stability of the laser output. The system is equipped with a phase-preserving pulse picking setup similar to the one described in~\cite{shestaev_high-power_2020}. To allow for a broadband measurement of the CEO noise up to a maximal available offset frequency ($f_\text{rep}/4 = 20$\,MHz), the original pulse repetition rate of 80\,MHz defined by the oscillator was, however, preserved.

The power amplifier is based on large mode area rod-type fibers with a mode field diameter of 65\,\textmu m and a length of 1\,m. The main amplifier contains 16 identical channels, which are coherently combined to provide up to 1.1\,kW of average optical power. By applying an amplitude and a phase mask in the front-end, an optical spectrum spanning over 11\,nm at a level of -10\,dB (Fig.~\ref{fig:spc_cpa}) and a pulse duration of 300\,fs was obtained. The measured autocorrelation function (ACF) with a FWHM of 420\,fs is shown in Fig.~\ref{fig:acf_cpa}. Further information about the system can be found in~\cite{haedrich_500w_2020}.

%FIG2: SPC & ACF
\begin{figure}[bp] %htbp
\centering
\captionsetup[subfigure]{labelformat=empty} % suppress (a) (b) subcaptions % requires package caption
  \subfloat[]{
    \includegraphics[width=0.78\linewidth]{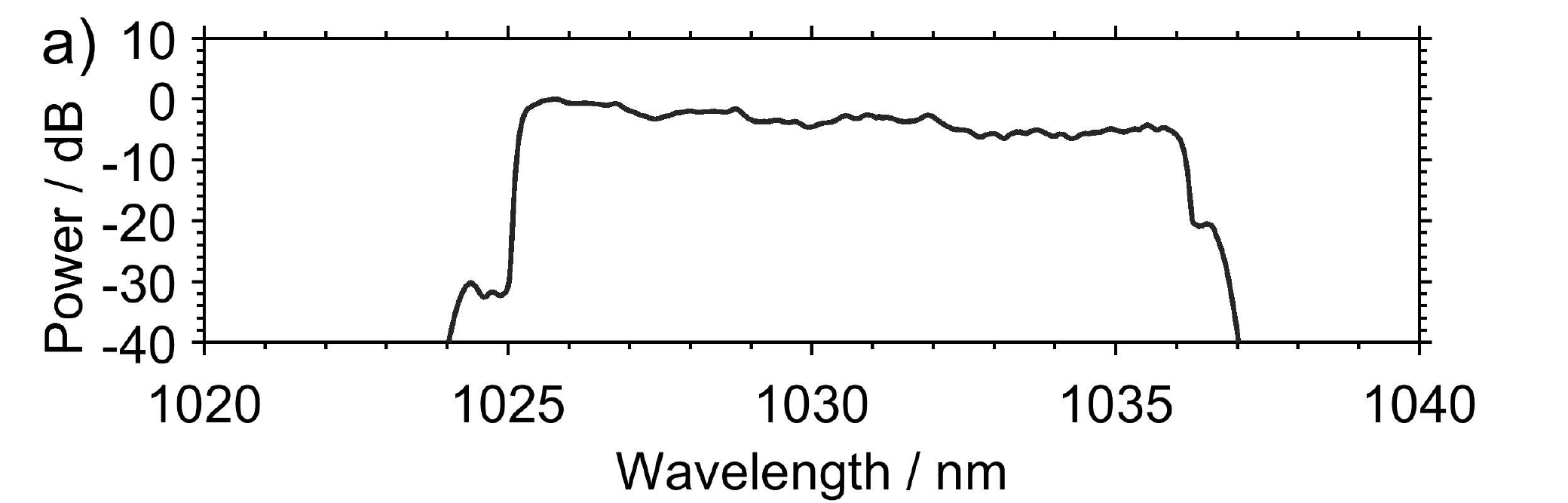} % , height=1.5cm
    \label{fig:spc_cpa}
  }
  %~
  \subfloat[]{
   \hspace{-0.07\linewidth}
    \includegraphics[width=0.25\linewidth]{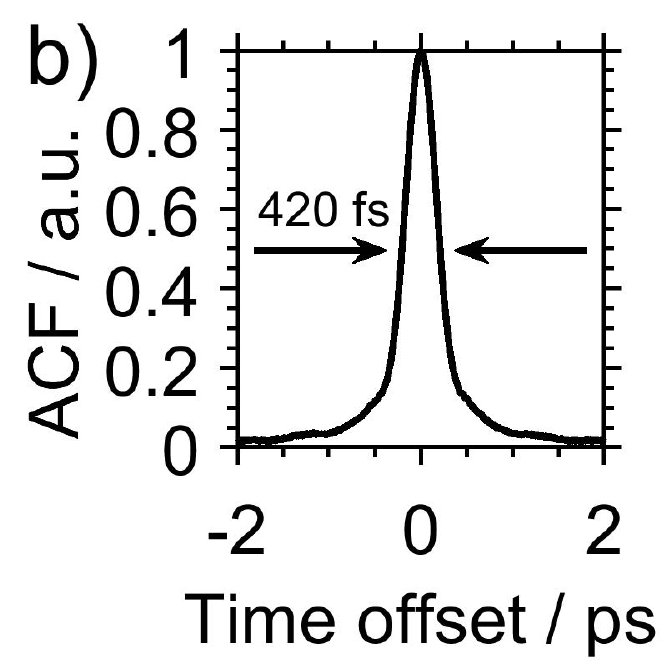} % , height=1.5cm
    \label{fig:acf_cpa}
  }
  \vspace{-4mm}
  \caption{Spectral and temporal characterization of the CPA system: a) Spectrum; b) Autocorrelation function (ACF).}
\label{fig:char_cpa}
\end{figure}

The CEO noise characterization unit only requires 1\,W of average power. Therefore, a significant power reduction had to be performed. To be able to cover the full average power range, i.e. operation of the system from 1 to all 16 channels, a combination of half-wave plates and thin-film polarizers was used. Due to the limited polarization extinction ratio they had to be operated away from the maximal suppression to ensure a linear polarization for the noise measurement setup.

%FIG3: SETUP_FINE
\begin{figure}[ht] % [htbp]
\centering
\captionsetup[subfigure]{labelformat=empty} % suppress (a) (b) subcaptions % requires package caption
  \subfloat[]{
    \includegraphics[width=1\linewidth]{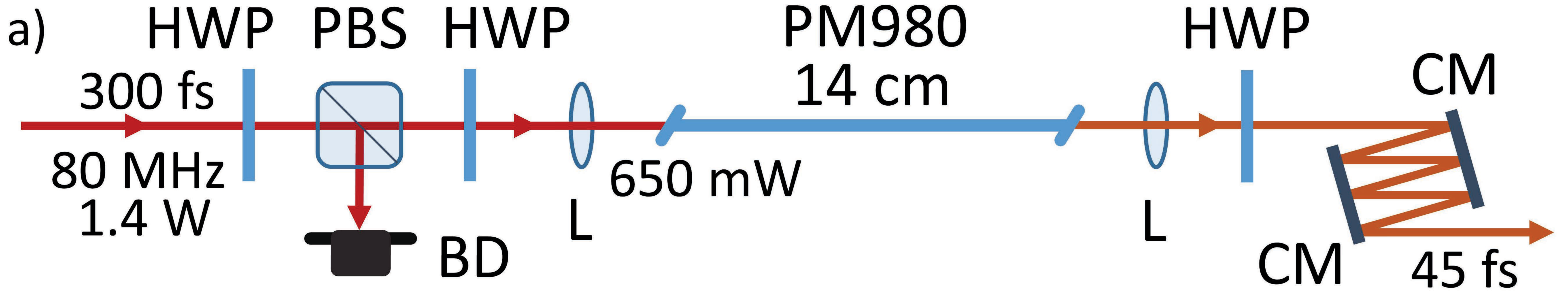}
    %\caption{}
    \label{fig:nlc}
  }
  %~%  fill line 
  \\ % new line
  \vspace{-8mm}
  \subfloat[]{
    \includegraphics[width=1\linewidth]{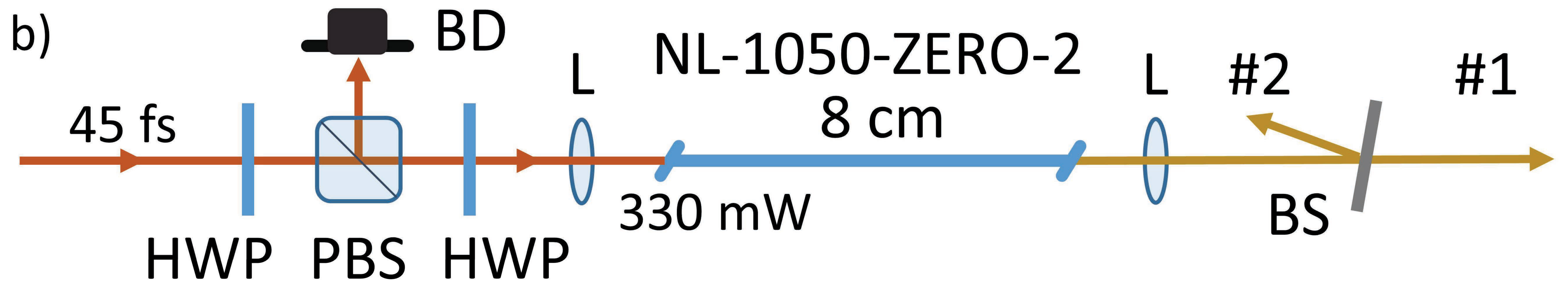}
    %\caption{}
    \label{fig:sc}
  }
  %\vspace{-6mm}
  \\
  \vspace{-7mm}
  \subfloat[]{
    \includegraphics[width=0.48\linewidth]{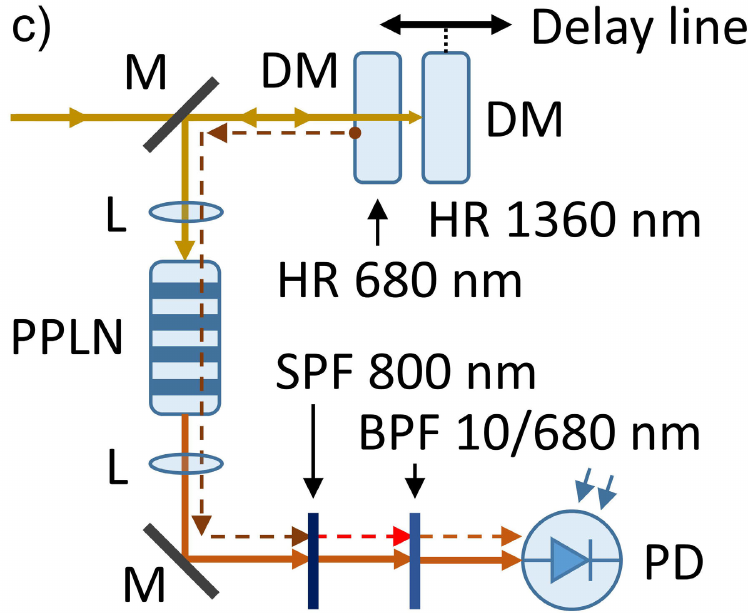}
    %\caption{}
    \label{fig:fto2f_setup}
  }
  %~ %  fill line 
  \subfloat[]{
    \includegraphics[width=0.5\linewidth]{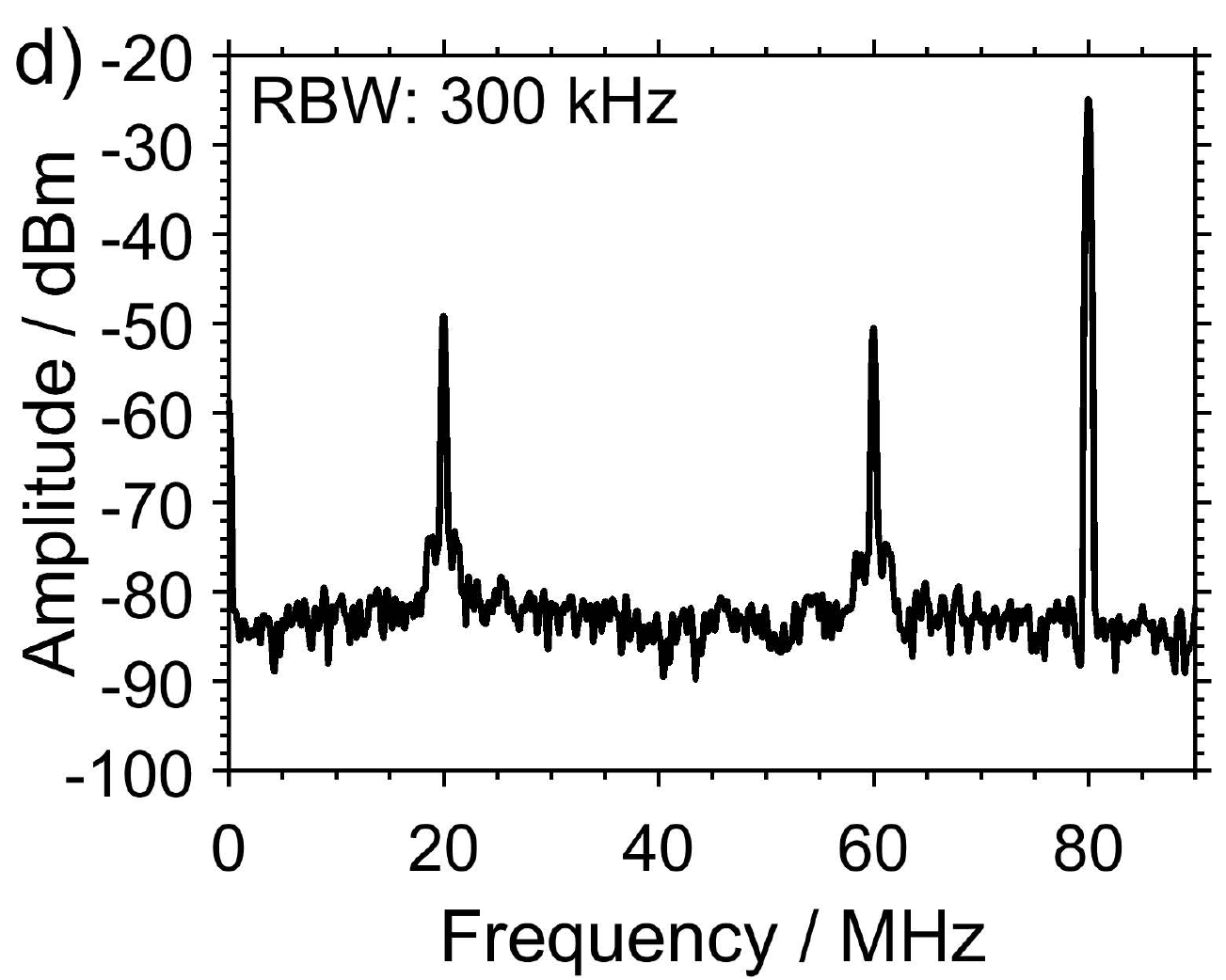}
    %\caption{}
    \label{fig:fto2f_beatnote}
  } 
  \\
  \vspace{-7mm}
  \subfloat[]{
    \includegraphics[width=0.78\linewidth]{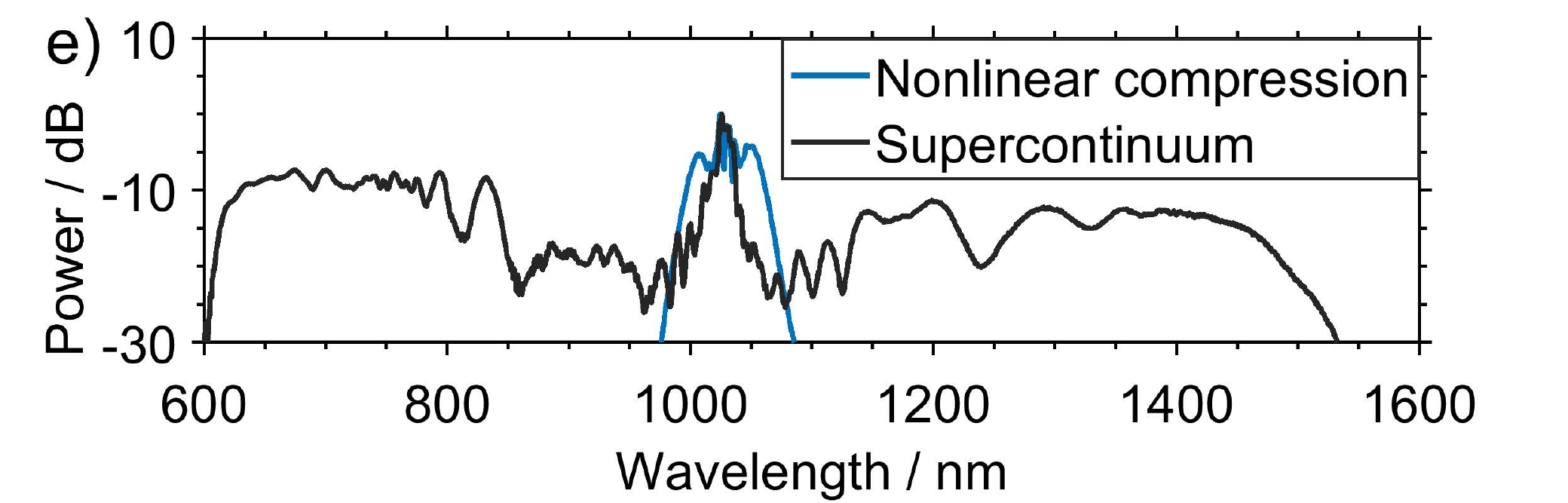} % , height=1.5cm
    \label{fig:spc_pre}
  }
  %~
  \subfloat[]{
   \hspace{-0.07\linewidth}
    \includegraphics[width=0.25\linewidth]{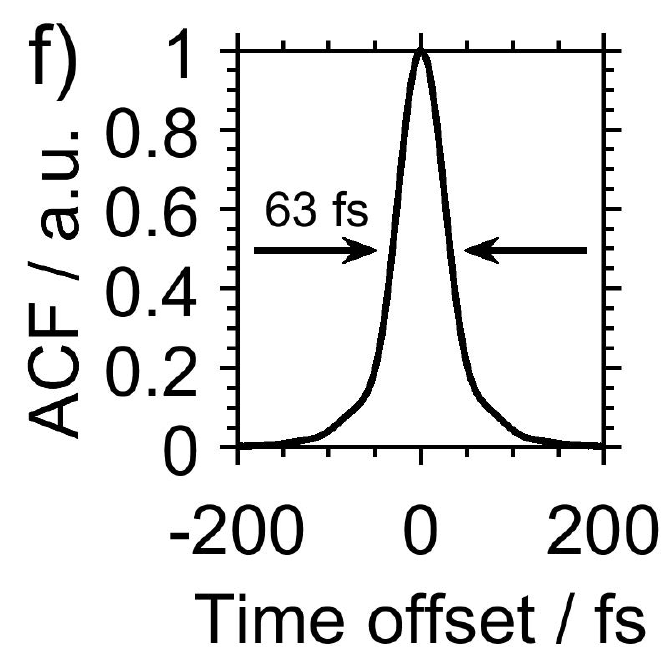} % , height=1.5cm
    \label{fig:acf_pre}
  }
  \vspace{-4mm}
\caption{CEO noise characterization setup: a) Nonlinear compression stage; b) Supercontinuum generation stage; c) f-to-2f interferometer; d) RF spectrum of the f-to-2f signal; e) Optical spectra at the nonlinear compression and supercontinuum generation stages; f) Autocorrelation measurement at the nonlinear compression stage; BD, beamdump; BPF, bandpass filter; BS, beamsplitter; CM, chirped mirror;  DM, dichroic mirror; HR, highly reflective; HWP, half-wave plate; L, lens; M, mirror; PD, photodiode; PBS, polarizing beamsplitter; PPLN, periodically poled lithium niobate; SPF, shortpass filter.}
\label{fig:comprsup}
\end{figure}
%}

In order to generate a coherent, octave-spanning supercontinuum~\cite{dudley_supercontinuum_2006}, the pulse duration is first reduced from 300\,fs to 45\,fs in a solid-core nonlinear compression stage (Fig.~\ref{fig:nlc}). It consists of a polarizing beamsplitter-based variable attenuator, a short piece of a passive single-mode fiber (PM980, 14\,cm) and a chirped mirror compressor (total GDD of -4900\,fs$^2$). The optical spectrum and the ACF measurement are shown in Fig.~\ref{fig:spc_pre} and~\ref{fig:acf_pre}, respectively. The supercontinuum generation stage (Fig.~\ref{fig:sc}) employs a short piece of an all-normal-dispersion photonic crystal fiber (NKT NL-1050-ZERO-2, 8\,cm)~\cite{heidt_pulse_2010}. The stage provides up to 330\,mW of white light spanning between 600 and 1500\,nm (Fig.~\ref{fig:spc_pre}), which is split equally between the two f-to-2f interferometers by a broadband beamsplitter.

The two f-to-2f interferometers (Fig.~\ref{fig:fto2f_setup}) are identical and follow the quasi-common-path scheme~\cite{jones_highly_2004,grebing_performance_2009}. The CEO beatnote is created by interference of a 680\,nm component of the supercontinuum with its frequency-doubled 1360\,nm component. Second harmonic generation is performed in a 1\,mm long periodically poled lithium niobate (PPLN) crystal. A set of spectral filters removes the light not contributing to generation of the beatnote, leaving 0.5 to 1.5\,mW of optical power for signal detection with a fast silicon photodiode. The RF spectrum of the photodetector output (Fig.~\ref{fig:fto2f_beatnote}) shows a high dynamic range of the beatnote of >30\,dB above the noise level at a resolution bandwidth (RBW) of 300\,kHz, which is enough for a tight phase lock~\cite{hartl_integrated_2005}.
The CEO locking electronics represents a phase-locked loop, which stabilizes the beatnote measured by the f-to-2f interferometer \#1 to 1/4th of the measured pulse repetition rate, in analogy to the stabilization system of the oscillator itself. The "slow loop" feedback signal is applied to an electro-optic modulator (EOM) located in the front-end of the CPA system, similar to~\cite{natile_cep-stable_2019}.

For generating the feedback signal we employed deeply modified commercial CEO locking electronics with external analog mixer and PI$^2$D controller. This unit provides a monitor port for the in-loop measurements and a feedback signal, which is applied to the EOM in the laser front-end via a homemade RF driver capable of delivering ±14\,V to a 50\,$\Omega$ load. The feedback loop has a 3\,dB bandwidth of 100\,kHz and a dynamic range of 7$\pi$ rad regarding to the optical path, which according to~\cite{natile_cep-stable_2019} corresponds to a CEP shift of 5.5\,rad.

%RESULTS
The CEO noise measurements were performed with a radio frequency spectrum analyzer equipped with a phase noise measurement option. The results (Fig.~\ref{fig:compare_fto2f},\ref{fig:ceo_lock},\ref{fig:ceo_cbc}) are represented as a log-log plot of the noise power spectral density (PSD) against the offset frequency. An offset frequency of 0 Hz would correspond to an absolute frequency of  $f_\text{ceo} = f_\text{rep}/4 = 20$\,MHz. Due to the symmetry of the beatnote and presence of its mirrored copy at $f = 3/4f_\text{rep} = 60$\,MHz, the entire information about the phase noise of the laser system is contained in an absolute frequency range of $f_\text{ceo}$ to $f_\text{rep}/2$ (20 to 40\,MHz), corresponding to an offset frequency range of 0\,Hz to 20\,MHz. The remaining part of the RF spectrum was suppressed by a 50\,MHz lowpass filter.

%FIG3: f-to-2f comparison
\begin{figure}[t]
\centering
\includegraphics[width=\linewidth]{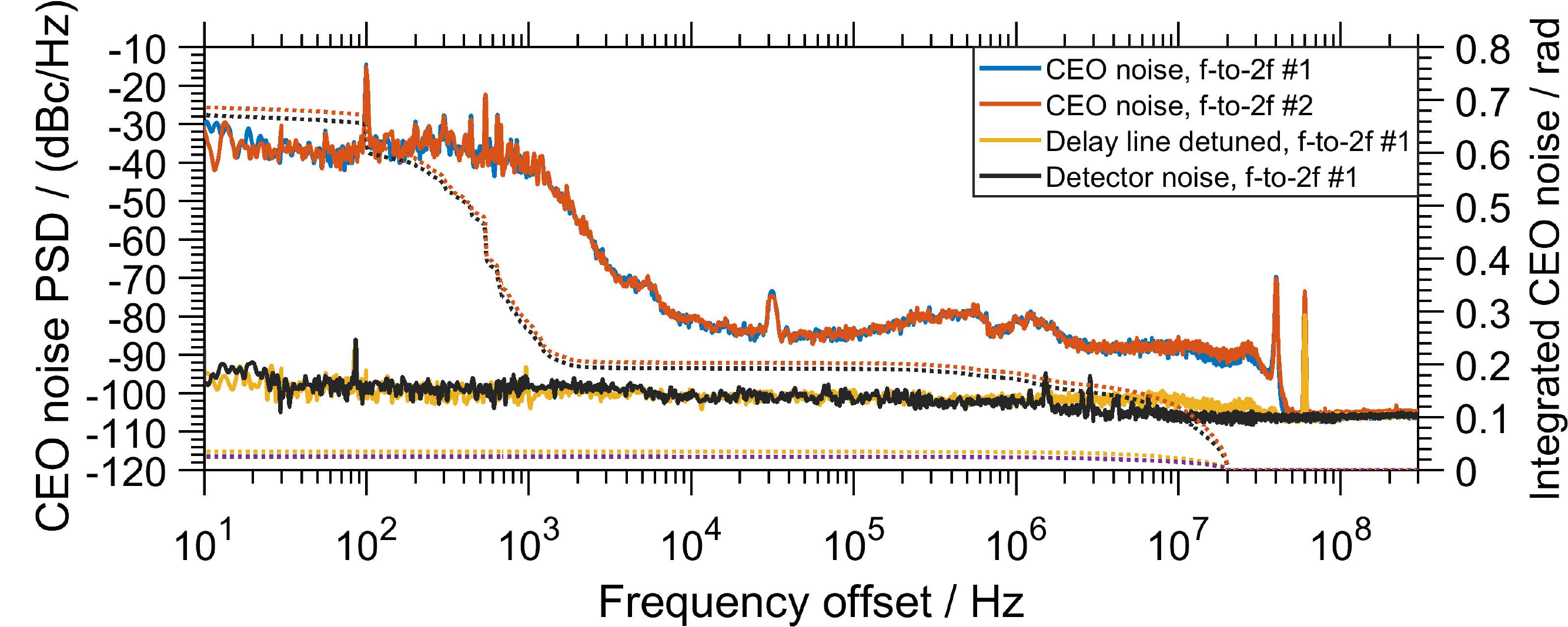} % , height=4cm
\caption{CEO noise spectrum measured by the f-to-2f interferometers.}
\label{fig:compare_fto2f}
\end{figure}

%FIG4: single channel
\begin{figure}[b] %[htbp]
\centering
\includegraphics[width=\linewidth]{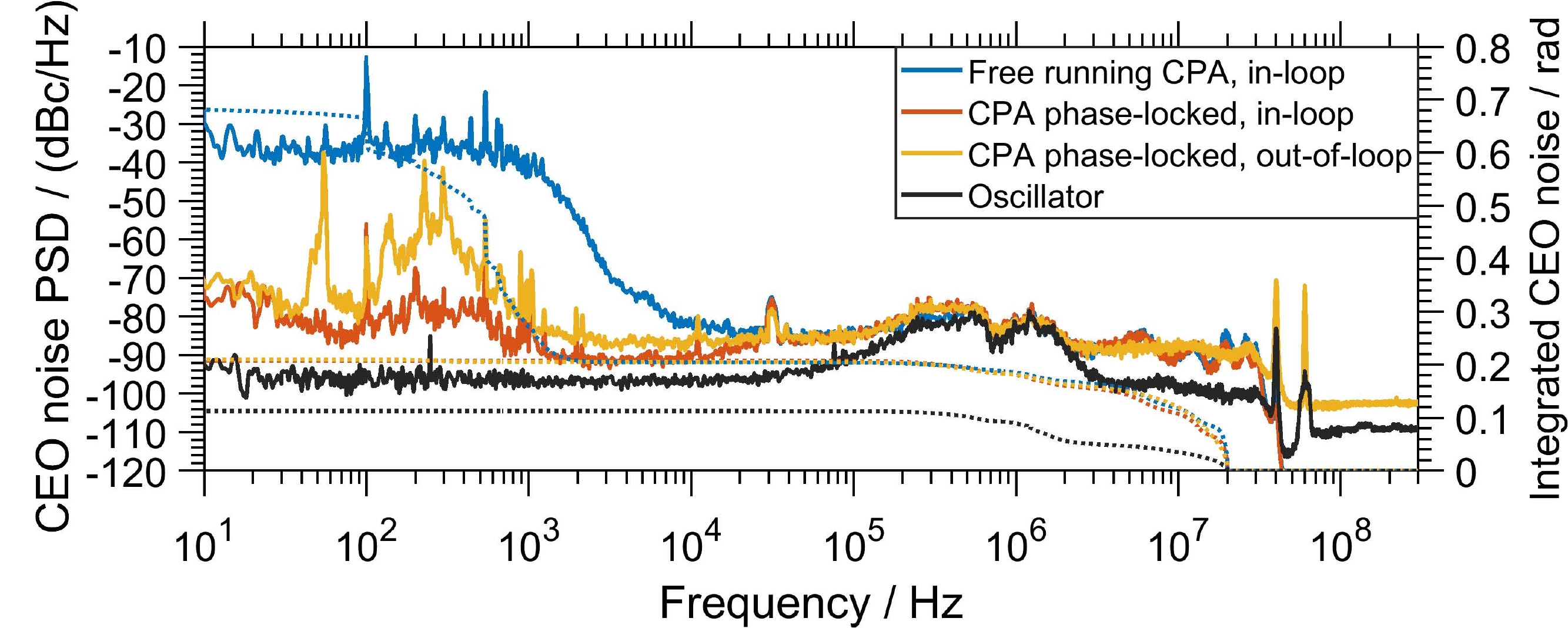} % , height=4cm
\caption{CEO noise of the CPA system with only one channel enabled in comparison to the oscillator.
}
\label{fig:ceo_lock}
\end{figure}

Phase noise measured at a ``free-running'' CPA system (only the oscillator phase-locked) by both f-to-2f interferometers is virtually identical (Fig.~\ref{fig:compare_fto2f}). This is important for a direct comparison of in-loop and out-of-loop measurements of a phase-locked CPA. It is also important to note, that the CEO noise in the entire frequency range of interest is above the noise floor of the measurement instrument. In addition, the noise level reduces to the instrument noise if the delay line of the f-to-2f interferometer is detuned. Since at the same time the average power at the photodetector remains constant, the observed excess broadband noise is not quantum-noise limited, but corresponds to the phase noise of the system.

As a first step, the CEO noise of the laser system with only one channel of the power amplifier enabled was characterized (Fig.~\ref{fig:ceo_lock}). A free-running CPA shows a relatively high integrated CEO noise of approximately 700\,mrad rms in an offset frequency range of 10\,Hz to 20\,MHz, however, a major part of it is accumulated at frequencies below 1\,kHz. Hence, the 100\,kHz bandwidth of the feedback loop is sufficient for noise correction.
With a feedback loop enabled, the in-loop CEO noise was suppressed to less than -70\,dBc/Hz, resulting in an integrated phase noise of only 220\,mrad rms at 80\,W of average optical power and a pulse repetition rate of 80\,MHz. The oscillations observed beyond 4\,MHz are due to transfer function ripple in the feedback electronics. The out-of-loop measurement reveals some remaining contributions in the acoustic frequency range, however, the integrated noise penalty in comparison to the in-loop measurement is negligible. 

The ultimate performance limit of the system is defined by the oscillator and amounts to 120\,mrad rms. The high power amplification system boosts the average power of the CEO stable oscillator by five orders of magnitude (from 10\,mW to 1\,kW), while increasing phase noise by only 100\,mrad. It is important to note, that the broadband, high-frequency noise creates the largest contribution to the integrated CEO noise of the system. Moreover, an increase of the noise beyond 2\,MHz at the CPA in comparison to the oscillator from -100\,dBc/Hz to -90\,dBc/Hz, which is still a very low level, effectively doubles the integrated CEO noise. On the one hand, this shows an extreme importance of characterizing the noise in the full available bandwidth of $1/4f_\text{rep}$ in the Fourier domain, or equivalently by single-shot methods without averaging over successive pulses in the time domain. On the other hand, investigation and suppression of high-frequency phase noise may provide room for further improvement.

To study the effect of the coherent beam combining on the noise performance of the system, we performed amplitude and CEO noise measurements with 1, 2, 4, 8 and 16 active channels in the main amplifier. For this experiment, every single amplifier channel was operated at a nominal average power of 80\,W, providing up to 1.1\,kW of average power with all 16 channels enabled. The total combining efficiency in this multilevel Hänsch-Couillaud setup~\cite{hansch_laser_1980,liang_coherent_2007} was over 95\%.

%FIG5: AN CBC
\begin{figure}[t] %[htbp]
\centering
\includegraphics[width=\linewidth]{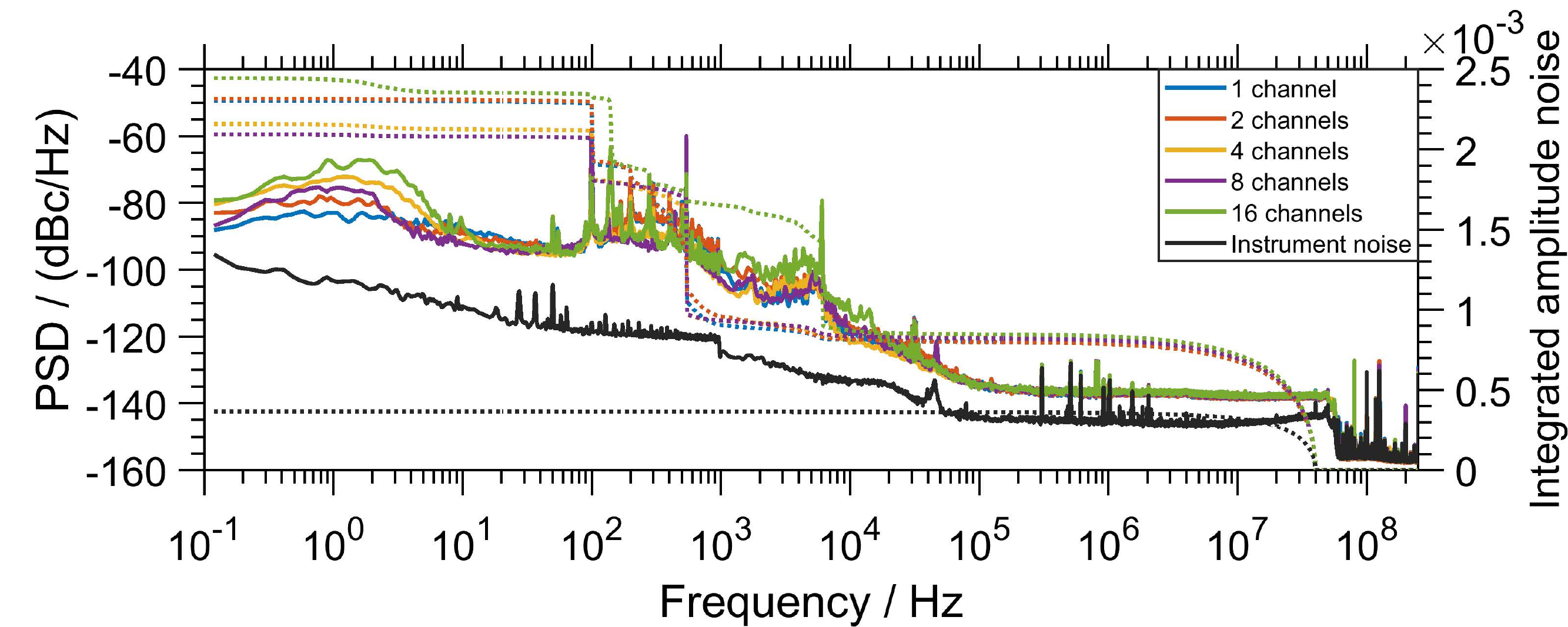} % , height=4cm
\caption{Amplitude noise of a coherently combined fiber laser system}
\label{fig:an_cbc}
\end{figure}

The amplitude noise of the coherently combined CPA system is shown in Fig.~\ref{fig:an_cbc}. The noise was measured with a setup consisting of an oscilloscope, a 50\,MHz low-pass filter and an amplified InGaAs photodiode. The general shape of the PSD curves is very similar for 1 to 8 active channels. Only when all the 16 channels are enabled and an output power of 1.1\,kW is reached, an increase of noise level between 1\,kHz and 20\,kHz can be observed. We attribute it to a higher thermal load of the system.
It does not, however, strongly affect the integrated relative intensity noise (RIN), which is mainly determined by low-frequency peaks below 10\,kHz and white noise above this frequency. Interestingly enough, the noise distribution between the individual peaks for each channel combination is different and shows no systematic behavior, but at the same time the total integrated noise for all the traces is similar. The rms RIN measured up to a half of the pulse repetition rate (0.1\,Hz to 40\,MHz) stays within 0.20\% to 0.25\% for all the configurations, which is an excellent result for a high-power laser system.

%FIG6: CEO CBC
\begin{figure}[t]
\centering
\includegraphics[width=\linewidth]{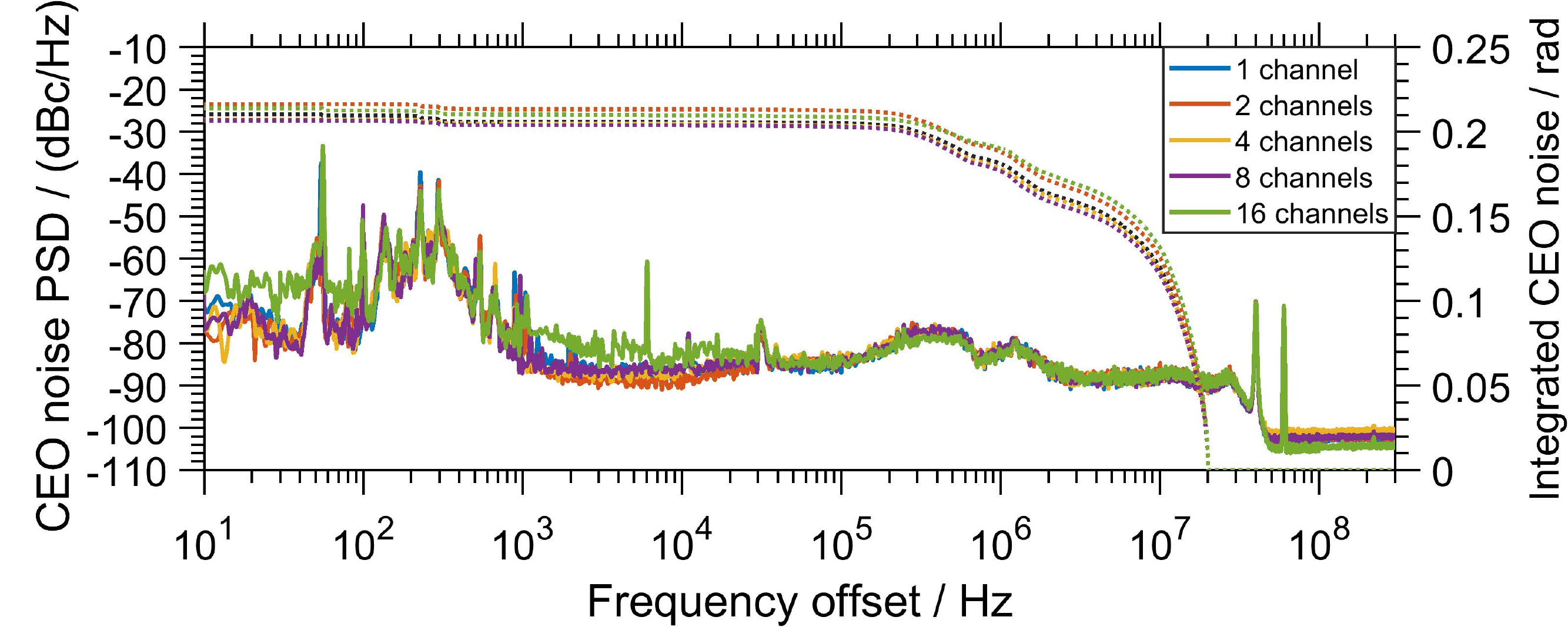} % , height=4cm
\caption{CEO noise of a coherently combined fiber laser system.}
\label{fig:ceo_cbc}
\end{figure}

The CEO noise of a coherently combined CPA system measured out-of-loop is shown in Fig.~\ref{fig:ceo_cbc}. Each trace is a power average of 4 measurements. The CEO noise of the CPA with 1 to 8 active channels is virtually identical, showing only a minor difference between 1\,kHz and 30\,kHz, which we attribute to fluctuation of the CEO beatnote power. Only at 1.1\,kW of average power (with all 16 channels enabled), two additional contributions appear: a 6\,kHz peak, which could be also observed in the amplitude noise, and a 1/f (flicker) noise. Their level does not exceed -60\,dBc/Hz, which is quite low, hence the contribution to the integrated noise remains negligible. For all the channel combinations, the integrated CEO noise remains fairly constant and amounts to only 220\,mrad rms (10\,Hz to 20\,MHz) -- a performance level which could previously be achieved by an optical parametric chirped-pulse amplification system with an average power of 53\,W~\cite{budriunas_53_2017}.

The long-term CEO stability of the system was limited by the dynamic range of the EOM. An uninterrupted phase lock was maintained for 5 to 15 minutes at an output power of up to 500\,W. At 1.1\,kW of output power the average duration of a CEO lock has reduced to the range of 30 to 60 seconds. Recently, the dynamic range of the actuator has been increased by an additional thermally controlled fiber spool, which allowed for several hours of an uninterrupted CEO-stable operation. 

To conclude, we have demonstrated the first CEO-stable, high power, coherently combined ytterbium-doped fiber CPA, which represents a basis of the ELI-ALPS HR2 laser system. Based on a robust fiber technology, the reported table-top laser system delivers a unique combination of an excellent CEO stability of 220\,mrad rms (10\,Hz to 20\,MHz), low amplitude noise of 0.25\% rms RIN (0.1\,Hz to 40\,MHz) at an average output power of more than 1\,kW. It is important to note, that the noise performance has been characterized in a bandwidth equivalent to a single-shot measurement. Both amplitude and CEO noise spectra are virtually independent of the number of the used channels, i.e. they exhibit a similar performance with 1, 2, 4 and 8 channels. With 16 channels and at the maximal output power a slight increase of noise level is observed, which does not significantly affect the integrated noise value.

In the nearest future, we plan to perform pulse energy scaling in this CPA to obtain 10\,mJ pulses with the same average power of 1\,kW and employ this system for generation of few-cycle, energetic, highly CEP-stable pulses at a repetition rate of 100\,kHz by means of nonlinear pulse compression. Some promising results in this direction have been already published in~\cite{nagy_generation_2019}. To provide further energy scaling of few-cycle pulses, the technology of divided pulse nonlinear compression can be employed~\cite{jacqmin_passive_2015}.

\medskip

\noindent\textbf{Funding.} 
H2020 European Research Council (835306, ``SALT''); Fraunhofer-Gesellschaft (Cluster of Excellence Advanced Photon Sources); European Regional Development Fund (GINOP-2.3.6-15-2015-00001, ``ELI-ALPS'').

\medskip

\noindent\textbf{Acknowledgements.} E. Shestaev acknowledges support by the German Research Foundation (DFG) within the International Research Training Group 2101.

\medskip

\noindent\textbf{Disclosures.} ES Active Fiber Systems (F); SH, NW, TE Active Fiber Systems (E); JL Active Fiber Systems (I, P).

%\section{References}

% Bibliography
%\bibliography{sample}
\bibliography{bibliography} 

% Full bibliography added automatically for Optics Letters submissions; the following line will simply be ignored if submitting to other journals.
% Note that this extra page will not count against page length
\bibliographyfullrefs{bibliography} 

\end{document}